\documentclass[graybox]{svmult}

\usepackage{mathptmx}       
\usepackage{helvet}         
\usepackage{courier}        
\usepackage{type1cm}        
\usepackage{makeidx}         
\usepackage{graphicx}        
\usepackage{multicol}        
\usepackage[bottom]{footmisc}
\usepackage[normalem]{ulem}

\makeindex             

\begin{document}

\title*{Parallelization of Kinetic Theory Simulations}
\author{Jim Howell, Wolfgang Bauer, Dirk Colbry, Rodney Pickett, Alec Staber, Irina Sagert, and Terrance Strother}
\authorrunning{J. Howell et al.}
\institute{
Wolfgang Bauer \at Department of Physics and Astronomy, Michigan State University, East Lansing, Michigan, 48824,
USA, \email{bauer@pa.msu.edu}
\and Dirk Colbry \at Institute for Cyber-Enabled Research, Michigan State University East Lansing, Michigan 48824, USA, \email{colbrydi@msu.edu}
\and Jim Howell \at Department of Computer Science and Engineering, Michigan State University, East Lansing, Michigan, 48824,
USA, \email{howell20@msu.edu}
\and Rodney Pickett \at Institute for Cyber-Enabled Research, Michigan State University East Lansing, Michigan 48824, USA, \email{picke1@mail.lcc.edu}
\and Irina Sagert \at Center for Exploration of Energy and Matter, Indiana University, 
Bloomington, IN 47408, USA, \email{isagert@indiana.edu}
\and Alec Staber \at Department of Physics and Astronomy, Michigan State University, East Lansing, Michigan, 48824,
USA, \email{staberal@msu.edu}
\and Terrance Strother \at XTD-6, Los Alamos National Laboratory, Los Alamos, New Mexico 87545, USA
}
%
%
\maketitle

\abstract{Numerical studies of shock waves in large scale systems via kinetic simulations with millions of 
particles are too computationally demanding to be processed in serial. In this work we focus on optimizing the parallel 
performance of a kinetic Monte Carlo code for astrophysical simulations such as core-collapse supernovae. 
Our goal is to attain  a flexible program that scales well with the architecture of modern supercomputers. 
This approach requires a hybrid model of programming that combines a message passing interface (MPI) 
with a multithreading model (OpenMP) in C++. We report on our approach to implement the hybrid design into 
the kinetic code and show first results which demonstrate a significant gain in performance when many processors 
are applied.}

\section{Computational Many-Body Methods}
\label{sec:intro}
During the last two or three decades computational approaches have become the `third leg' of science, complementing 
the more traditional theoretical and experimental/observational approaches.  In particular in studies of dynamical systems
additional insights can be obtained from large-scale computer simulations. Among the most popular approaches are
molecular dynamics simulations with explicit two-body interactions, kinetic theory approaches solving Boltzmann like 
equations with mean field and two-body collision terms, and hydrodynamics studies, which are based on short mean-free-path
assumptions.

In nuclear physics, in particular, the field of heavy ion reactions at intermediate and high energies has been studied extensively
on the basis of hydrodynamics \cite{SMG74,SMG80,Bou09,LP08,SH09,SJG11,Nov13}, kinetic theory 
\cite{Won82,Ber84,Kru85,Bau86,SG86,Ber88,Sch89,Gon91,Wan91,LKB98,Lin05,LCK08}, and molecular dynamics 
\cite{Wil77,BP77,Aic87,Fel90,Aic91,FS00} methods. Since the physical phenomena in heavy ion collisions require spatial
resolution of on the order of $10^{-15}$ m and time resolution of on the order of $10^{-24}$ s, direct observations are not 
possible, and careful modeling of reaction dynamics is essential for any meaningful comparison with experimental observables.
Thus methods like the ones just listed are indispensable tools for heavy ion physics. In particular, one can show that several
of these approaches have identical results in certain analytic test cases, which increases our degree of confidence in the
simulation results \cite{Bou09,Sag12,Sag13}.

During the last few years our group has begun to apply the heavy ion modeling expertise to the problem of supernova explosion.
The preferred tool to study these explosions has been hydrodynamics simulations for the baryons, coupled to Boltzmann transport
for the neutrinos \cite{Wil85,Her94,FH00,Mez01,FW02,TBP03,Heg03,Bur03}. However, it can be shown that kinetic theory based
models can treat the baryon and neutrino transport on the same footing within the framework of a coupled transport theory
\cite{Bau04}, similar to the coupled transport relativistic transport theory for baryons, mesons, and resonances for heavy ion
collisions \cite{Wan91}.  This enables us to conduct fully relativistic 3+3 dimensional studies of supernova core collapses
\cite{Bol02,SB07,SB09, Str09,SB10}.  In order for these studies to be able to provide reliable predictions that can be compared 
to observables, however, one has to utilize a very large number of test particles, on the order of $10^8$. This, in turn, requires the use of
modern multi-processor environments with effective algorithms for parallelization.  The present manuscript describes such algorithms
and our implementation.

\section{Initial Simulation Setup}
\label{sec:1}
Our kinetic simulation code models hydrodynamic shock waves by tracking the collisions of
millions of particles over a set number of time steps. Each collision is determined by the distance of
closest approach method. Hereby, each particle looks for its nearest collision partner at every time
step. The naive approach to this type of collision detection problem is to compare the position and
velocity of every particle to every other particle in the simulations to find a collision partner. The 
computational complexity of such an algorithm scales with the total number of particles $N$ as $O(N^2)$. 
For smaller scale problems, this algorithm is sufficient. However, for larger scale problems the amount of 
computation time required to run such a simulation is unreasonable. Achieving a minimal processing time 
is a common challenge for computationally demanding computer models. Parallelization is a key ingredient in 
these simulations as it divides up the workload among multiple processors. The algorithm we chose 
is conceptually simple. The simulation space is divided up into a grid of cells or so-called bins. Hereby, each
bin is represented by a pointer. The grid is thereby setup as a multidimensional array of pointers. The number 
of bins in the $x$, $y$, and $z$ dimensions and the size of the simulation space in each dimension are set by the user. 
The characteristic length scale of the system corresponds to the width of a bin, $\Delta x$, and is therefore determined 
from the size of simulation space and the number of bins. The mean free path must be set significantly smaller 
than the characteristic length scale to ensure the simulation is fluid-like. We use an adaptive time 
step size $\Delta t = \Delta x / v_{max}$, where $v_{max}$ is the maximum speed of the particles. With that, we 
can ensure that no particle can escape its scattering neighborhood within a single time step. Furthermore, 
if the difference between the average particle velocity and the maximum velocity is sufficiently small, this 
modification guarantees that the time steps will not be too small. Each particle is an instance of a class 
that contains a particle's velocity, position, mass, effective radius, and two particle pointers. One pointer 
is used to  identify the collision partners, while the other is to connect particles that are located within the 
same bin via a linked list.

At the beginning of the simulation, all particles are initialized within a box. The boundary
conditions of the box can be set as reflective, absorptive, or periodic and depend, like the exact 
initialization of the particles, on the particular test being run. Particles are sorted into their respective bins by
appending them to a linked list based upon their initialized position. Once all of the particles have
been sorted, output data is processed and printed to a file.

When output is complete, the binary collision detection phase of the program starts. This is the
most computationally intensive part of the program, and we make use of the cell linked list
structure to achieve greater computational efficiency. Our program loops through all bins, with 
every particle in a bin being compared to all of the other particles in the neighboring bins 
(26 bins in three dimensions and 8 bins in two dimensions). A particle is considered as a 
collision partner if the following criteria are met: 
\newline
The collision partners 
\begin{itemize}
\item are located within each other's collision neighborhood, 
\item are within each other's effective radii,
\item and reach their distance of closest approach within the next time step.
\end{itemize}
If a potential collision partner is found, the pointer of the searching particle is set to point to this
new collision partner. If a closer collision partner is found later in the search, then that pointer is 
updated accordingly to the new particle. In this way, each particle finds its closest collision
partner without searching through the entire simulation space. The computational efficiency of 
this algorithm scales as
\begin{eqnarray}
O(B\times N_{bin}^2/P), 
\label{eff}
\end{eqnarray}
where $B$ is the number of bins, $N_{bin}$ is the maximum number of particles per bin, and
$P$ is the number of processors. As opposed to scaling exponentially with the total number of particles,
our code scales exponentially with the number of particles per bin, which can be controlled by using a
large number of bins when there many particles. Note that the binary collision
detection step does not update any particle properties other than the pointers to collision partners.

The number of processors $P$ resides in the denominator of eq.(\ref{eff}) due to the parallelization of 
the binary collision detection phase. We use OpenMP to create a new slave thread for each processor in 
addition to the processor running the master thread. Since OpenMP uses a shared memory model, every 
thread can access the same set of pointers and variables. Each thread searches approximately an equal 
number of bins for scattering partners. No two threads are allowed to compute the same bin and its collision 
neighborhood. This is a simple approach for keeping the workload of all of the processors roughly the same. 
However, in tests with shock fronts, some bins can have drastically higher particle densities and velocities 
than others. These bins can take significantly more computational time than bins with lower particle densities, which 
creates a load imbalance. When a load imbalance occurs, some processors complete their computation before 
others and have to wait for the other processors to finish before proceeding. A more 
sophisticated approach to dividing up the workload, for example an adaptive bin size, will be 
considered in future versions of this code. At the end of the binary collision step, all of the slave threads 
share their information with the master thread and subsequently close. The code beyond this point executes 
in serial on the master thread. 

The final stage of the simulation updates particles' positions and velocities. Two particles only collide if they 
have both found each other as a collision partner in the binary collision detection phase. If this occurs, the 
particles' positions and velocities are calculated accordingly, and boundary conditions are applied to particles that
come in contact with the walls of the simulation space. The program will then loop back to the part of the code that
sorts the particles into their bins, and then repeat the steps described above. 

\section{Algorithm Optimization}
\label{sec:2}

\begin{figure}
\sidecaption
\includegraphics[width=0.6\textwidth]{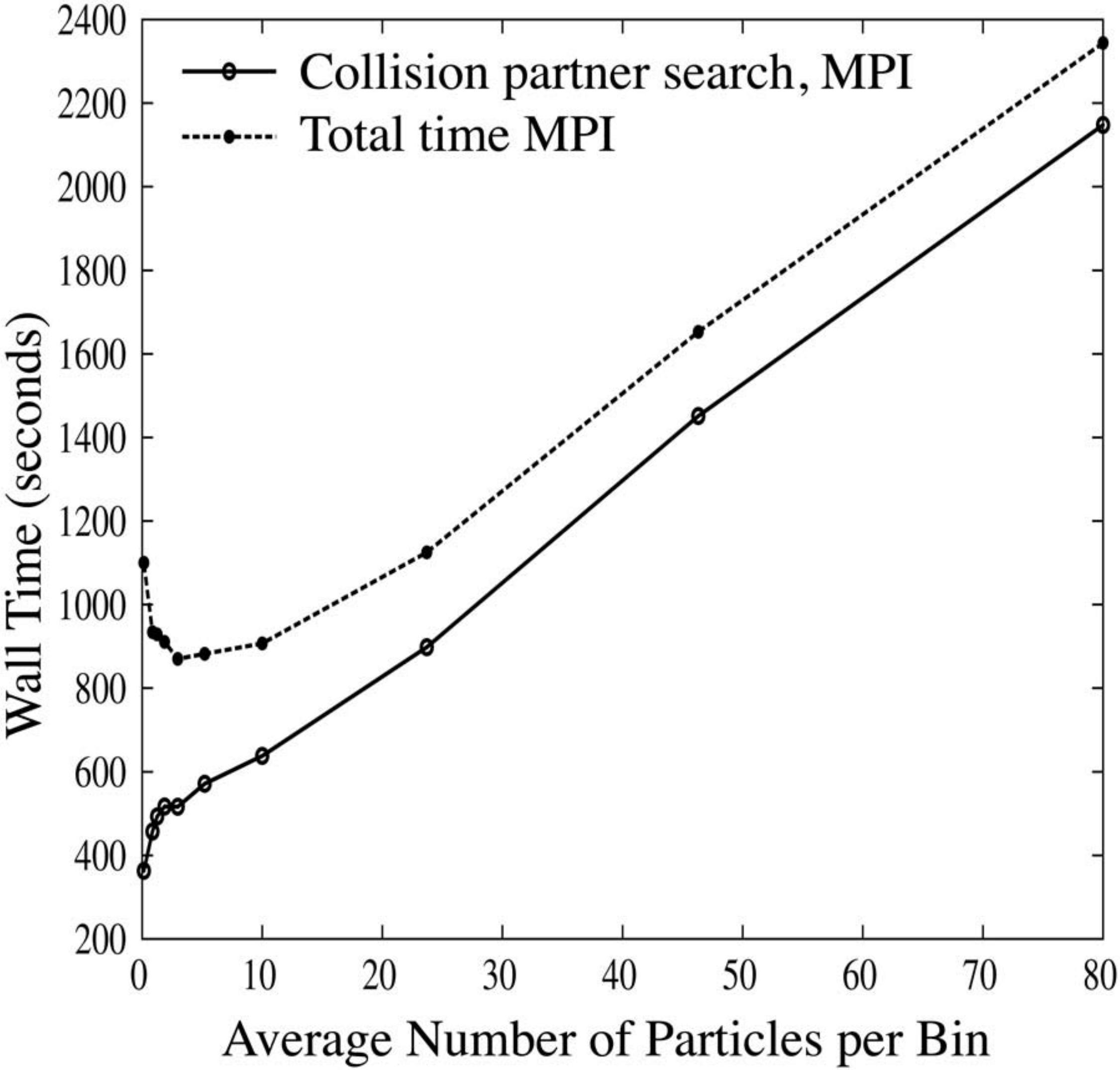}
\caption{Timing study for the wall time of our initial simulation as a function of the average number of particles per bin.  Each test was initialized with $10^8$ particles in three dimensions, with all of the particles having the same initial speed, random velocity directions, and random positions. The number of time steps taken in each test ranges from 20 to 206. The tests with a lower average number of particles per bin take more time steps to maintain the same elapsed simulation time. Each test produced a total of 40 output files.}
\label{particleperbintiming}
\end{figure}

Timing studies show that our algorithm can be optimized by using a low average number of
particles per bin, with some caveats. Fig.\ref{particleperbintiming} illustrates that the relationship between 
computation time and number of particles per bin is nearly linear when the average number of particles per 
bin is significantly greater than one.  This indicates that decreasing the number of particles per bin can 
significantly improve computation time. However, at roughly three particles per bin, the total wall time reaches 
a minimum, and the correlation between the total simulation time and binary collision  detection time begins 
to diverge. Below this value, the total computation time starts to increase and the binary collision detection 
computation time starts to decrease. Further testing is needed to understand the cause of this divergence, 
and how using a low number of particles per bin could effect shock fronts.

One limitation to this optimization method is high random access memory usage. Our code
uses a multidimensional array of pointers to sort the particles. Increasing the number of bins increases
the size of that array. In 2D simulations, this memory increase is generally not large
enough to be a significant limitation, but for 3D simulations this can be a constraint for
simulations with $\gg 10^8$ particles and a correspondingly large number of bins. 
Nonetheless, modern supercomputers can usually handle this amount of memory consumption.

Furthermore, changing the number of bins affects the time step size of the simulation. Our code has an 
adaptive time step that is proportional to $\Delta x$. For a square or cubic simulation space, 
increasing the number of bins in the simulation decreases the time step size, and the simulation will need
approximately twice as many time steps to reach the same simulation time. To negate this potential
drawback, a time step counter ensures that output is only generated at a specified time interval. 
This allows a user to maintain the same amount of output regardless of the time step
size. To control for the adaptive time step issue, all of the tests were run with the same amount of total 
output and elapsed simulation time.

\section{Hybrid Simulations}
\label{sec:3}
Implementing MPI is generally more difficult than implementing OpenMP. OpenMP uses
straightforward preprocessor directives called pragmas that signal to the compiler when a section of
code can be parallelized. The creation, destruction, and communication of threads is established
automatically by compiler at these points, and most of the complex details are hidden from the user.
One can implement more specific control over threads with work sharing constructs and through the
manipulation of environmental variables to optimize performance. One advantage of using OpenMP is
that a program can still use the same compilers after parallel pragmas have been added. This allows for
incremental parallelization, without the need for creating separate files for an OpenMP version of the
code. OpenMP pragmas ordinarily are placed before the most computationally intensive loops of the
program, and the remainder of the code executes serially. One limitation of this method of parallelization is 
that simulations cannot run on multiple nodes simultaneously, because nodes use a distributed memory system.

In contrast, using MPI requires a better understanding of parallelization and the code that it is
being applied to. MPI uses a standardized set of library routines to pass bits of information and
instructions from one processor to another. Often, this type of parallelization involves determining how
to break a large problem into smaller pieces that share information between each other with MPI
function calls. This process is referred to as domain decomposition. Domain decomposition can
require a significant amount of code rewriting, particularly if it is being considered after a code already
been developed. An abstraction called ghost cells are frequently used in conjunction with domain decomposition. 
Ghost cells are copies of neighboring cells from different nodes, which allow each node to have access to the 
information from other nodes.  

The benefits of using MPI are that it allows: for finer control of parallelization, for
parallelization of the entire code, and a parallelized code to run on a distributed memory system. By implementing 
a hybrid of both OpenMP and MPI, we can gain the benefits of both types of parallelization, while 
mitigating some of the drawbacks of each. There are various ways to structure a hybrid code. A standard 
method that we use in our code is to create one MPI thread per node and use MPI routines to 
communicate between each node. OpenMP is used to create additional threads on each node in 
the binary collision detection phase.

\begin{figure}
\sidecaption
\includegraphics[width=0.62\textwidth]{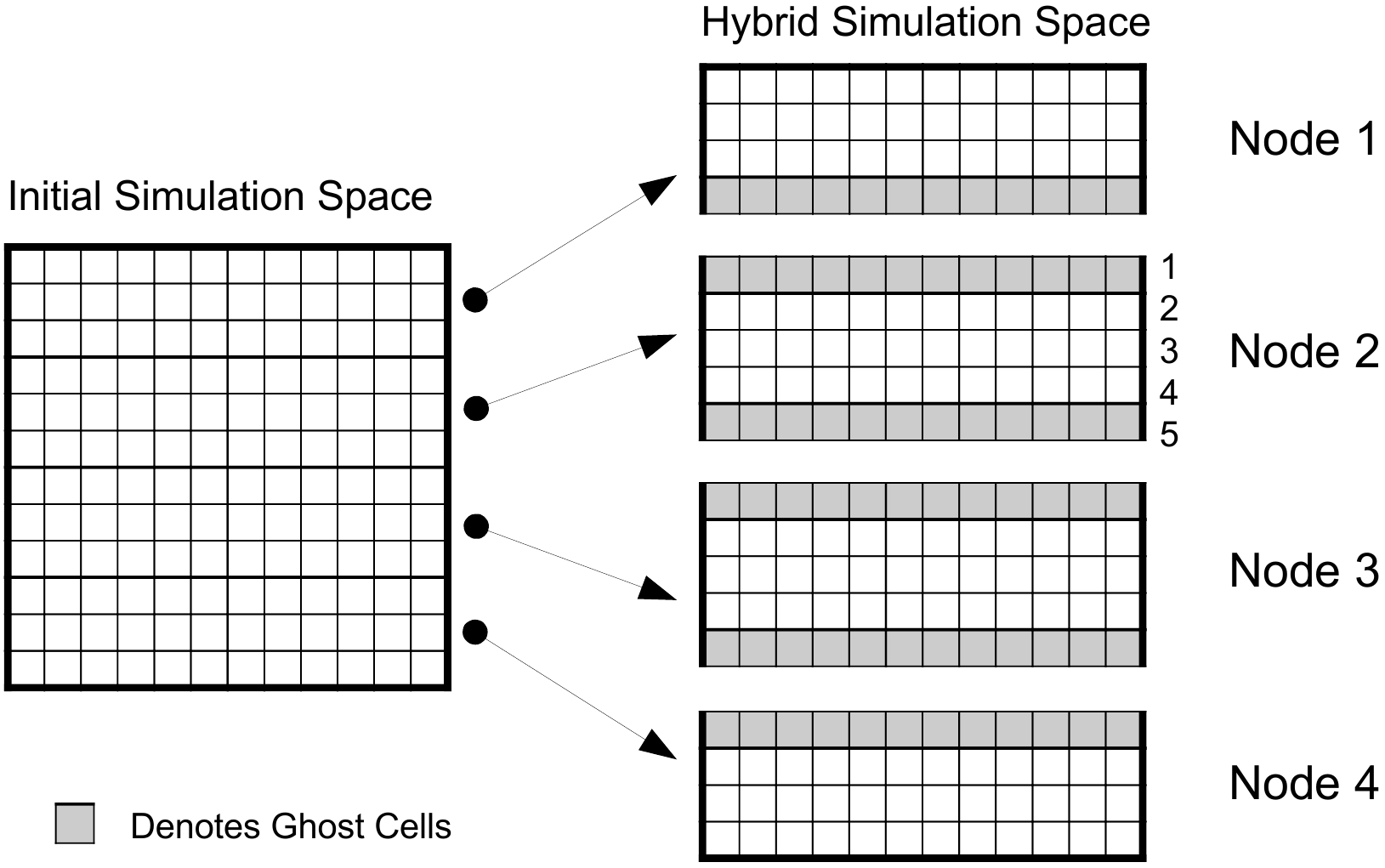}
\caption{Initial and hybrid simulation spaces.}
\label{diagram}
\end{figure}

For domain decomposition, we divide the simulation space by the number of nodes in use. The
number of rows in our simulation must be evenly divisible by the number of nodes to split up the
simulation space into equal parts. For example, if a simulation with 12 rows is running on four nodes,
then the grid of the simulation space is split up into four domains of three rows each. Each domain is
executed on a different node. In this case, the two nodes designated to handle the middle of the
simulation space both have two neighboring domains: one domain above and one domain below. The
middle domains have one empty row added above, and one empty row added below their normal
simulation space. These extra rows are used to store a row of ghost cells that they receive from their
respective neighboring domains. A node designated to run the bottom part or the top part of the
simulation space is a special case. In this case, a node has only one neighboring domain, because it lies
on the boundary of the simulation space. In practice, there can be an arbitrary number of middle
domains, but there are always two special case domains on the boundaries. Fig.\ref{diagram} demonstrates this example.

We have added four stages of communication between nodes to our simulation. The first stage
of communication transfers rows of ghost cells from neighboring nodes. The second phase syncs the
adaptive time step of all the nodes. The third phase sends the collision information of the ghost
particles found in the collision detection phase. The lasts phase of communication transfers the real
particles between nodes.

In the first stage of communication, we count the number of ghost particles that will be sent
from each node and send that information to the neighboring nodes. This information allows each node
to know how many ghost particles it will receive. In the example with four nodes
described above, each middle domain has a total five rows. The first and the fifth rows are the ghost
rows used for storing information received from other nodes. The second and the fourth rows contain
the particle information that is sent to neighboring nodes. These particles are packed into a contiguous
memory storage, sent individually with an MPI send function, and received with an MPI receive
function. For the middle nodes, all of the particles in the second row are sent to the node above in the
simulation space and copied into that node's fifth row. All of the particles in the fourth row are sent to
the node below in the simulation space and copied into that node's first row. For the special case nodes,
the top node sends the particle information from its fourth row to the first row of the node below, and
the bottom node sends particle information from its second row to the fifth row of the node above.
Once the first communication phase is complete, all of the nodes' ghost rows are filled with copies of
particles from the neighboring rows of other nodes. This enables the binary collision detection phase to
find all of the potential collision partners for particles in the second and fourth rows.

The second stage of communication is brief but important. Since all of the nodes are essentially
running separate simulations that intermittently communicate with their neighbors, each node can
develop a different maximum particle velocity. Our adaptive time step is derived from the maximum
particle velocity, so each domain can have different sized time steps if they are not synced. By using an
MPI reduce function, all of the nodes can acquire a global maximum particle velocity, and use that
information to set their adaptive time steps to the same value.

The third communication phase sends collision partner information found in the binary collision
detection step from particles in the second and fourth rows to the ghost particle copies of themselves.
The pointers must be converted into particle ID integers, since pointers are not meaningful when
transferred to other nodes. In our four node example, if a real particle in the fourth row has found a
ghost particle as a collision partner in the fifth row, then that particle's collision information will be sent
to the first row of the node below. The pointers of the corresponding ghost particle will be updated with
this information. This process repeats until all of the particles in the second and fourth rows with ghost
particle collision partners have sent their collision information. Without this information, there is no 
way of determining if the ghost particle in the fifth row has also found the real
particle in the fourth row as a collision partner.

The lasts phase of communication occurs after the collisions have been calculated. Once the
particles' positions and velocities have been updated, and boundary conditions have been applied, the
code counts the number of particles with coordinates outside of their respective domain spaces.
Particles are removed from the domains that they are leaving and added to the domain that they are
entering. In our four node example, if a real particle lies in the first row of a middle node, it will be
transferred to fourth row the node above, and if a real particle lies in the fifth row of a middle node, it
will be transferred to the second node of the node below. Each node has its own coordinate system, so
the particles' positions are transformed accordingly. The program either stops after this point or loops
back to the part of the code that sorts particles into their bins, depending on the number of time steps.

\begin{figure}
\sidecaption
\includegraphics[width=0.62\textwidth]{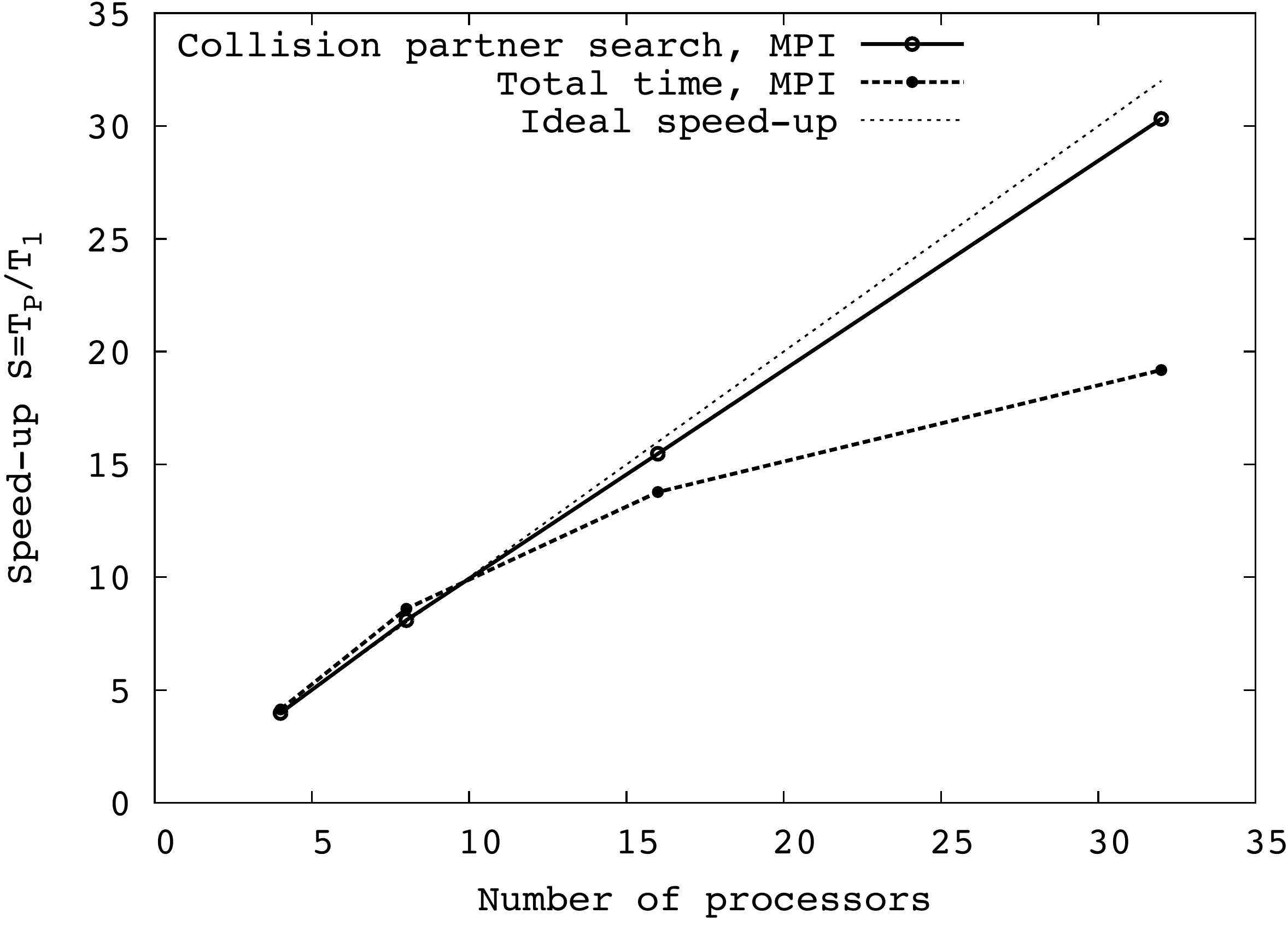}
\caption{Timing test for the speed-up of our hybrid implementation as a function of the number
of processors (4, 8, 16, and 32 nodes with one processor per node) used in the calculation. The dotted line shows the theoretical maximum speedup
possible. Each test was initialized with 40,000,000 particles in two dimensions, with all of the particles having the same initial speed, random velocity directions, and random positions. Each test produced a total of two output files per node.}
\label{mpitiming}
\end{figure}

The advantage of our hybrid code is that each node runs almost entirely in parallel. The
drawback is that there is significant overhead and processing time required to send the data of millions
of particles between nodes. Also, each node produces a separate output file, which are combined after
the simulation is complete. Yet, as the empirical timing studies in Fig.\ref{mpitiming} show, the binary collision detection phase of hybrid code scales well with an increasing number of processors. Furthermore, this setup should allow the code to run with hundreds or thousands of processors, and should 
greatly reduce job scheduling times. 

\section{Outlook}
\label{sec:4}

So far our hybrid code has only been developed in two dimensions for an initialization in which all of
the particles have random positions, random velocity vector directions, and reflective boundary
conditions. This scenario is ideal for parallelization since each processor has approximately the same
workload. The initializations for the supernova simulations are more complex and will require more
careful consideration for implementation in the hybrid version. The supernova tests with shock fronts
will probably not scale as well. One solution for this issue is to use dynamic bins sizes on the level of
the OpenMP implementation, and dynamic domain sizes on the level of the MPI implementation.
Areas of higher density would have smaller bins sizes and domain sizes than areas of lower density.

\paragraph{Acknowledgements}
The authors would like to thank the Blue Water Undergraduate Petascale Education Program and Shodor for their financial and 
educational support. Furthermore, this work used the Extreme Science and Engineering Discovery Environment (XSEDE), which is supported by 
National Science Foundation grant number OCI-1053575. I.S. is thankful to the Alexander von Humboldt foundation and acknowledges 
the support of the High Performance Computer Center and the Institute for Cyber-Enabled Research at Michigan State University.

\end{document}